\documentclass[aps,pra,twocolumn,superscriptaddress,10pt,longbibliography]{revtex4-2} 
\usepackage{graphicx}
\usepackage{dcolumn}
\usepackage{bm}

\usepackage[utf8]{inputenc}
\usepackage{amssymb}   
\usepackage{amsmath}   
\usepackage{braket}    
\usepackage{subfig}    
\usepackage{comment}   
\usepackage{float}	   
\usepackage{ctable}    
\usepackage{multirow}
\usepackage{color}
\usepackage{fancyhdr}
\usepackage{amsthm}
\usepackage{tabularx}
\usepackage{setspace}
\usepackage[export]{adjustbox}
\usepackage{stfloats}
 \usepackage{tabu}

\usepackage{fancyhdr}
\begin{document}
\fancyhead[R]{\ifnum\value{page}<2\relax\else\thepage\fi}

\preprint{APS/123-QED}

\title{Enhanced Hong-Ou-Mandel Manifolds and figures of merit\\ for linear chains of identical micro-ring resonators}
\author{Peter L. Kaulfuss}\email{corresponding author: plk3729@rit.edu}
\affiliation{Air Force Research Laboratory, Information Directorate, 525 Brooks Rd, Rome, NY, 13411, USA}
\affiliation{School of Physics and Astronomy, Rochester Institute of Technology, 85 Lomb Memorial Drive, Rochester, New York 14623, USA}
\author{Paul M. Alsing}
\affiliation{Air Force Research Laboratory, Information Directorate, 525 Brooks Rd, Rome, NY, 13411, USA}
\author{A. Matthew Smith}
\affiliation{Air Force Research Laboratory, Information Directorate, 525 Brooks Rd, Rome, NY, 13411, USA}
\author{Joseph Monteleone III}
\affiliation{School of Physics and Astronomy, Rochester Institute of Technology, 85 Lomb Memorial Drive, Rochester, New York 14623, USA}
\author{Edwin E. Hach III}
\affiliation{School of Physics and Astronomy, Rochester Institute of Technology, 85 Lomb Memorial Drive, Rochester, New York 14623, USA}

\date{\today}

\begin{abstract}
We present an exact analytic expression for the Hong-Ou-Mandel manifold (HOMM) for any number of identical Micro-Ring Resonators (MRRs) in a linear chain. The HOMM is the higher-dimensional manifold of solutions where the HOM effect is obtained; for the types of MRR systems investigated in this work, it is a one dimensional curve. We investigate the extreme stability of this HOMM, showing that the HOM effect in linear arrays of MRRs is highly robust. We further use this expression to derive three figures of merit for the HOMM of linear chains of MRRs: 
the minimum MRR coupling value ($\tau_{c}$), 
the curvature of the 1-dimensional HOMM ($\bar{\xi}_N$), 
and 
the 5\% tolerance in the probability for complete destructive interference to occur on the HOMM ($\delta\tau_{N}$). We promote these metrics to characterize the pros and cons of various linear chains of MRRs and inform design and fabrication.

\end{abstract}

\maketitle
\thispagestyle{fancy}


\section{Introduction}
The Hong-Ou-Mandel (HOM) Effect, originally discovered in 1987 as the essential phenomenon underlying then groundbreaking sub-picosecond time delay measurements \cite{HOM:1987,Fearn:1987,Rarity:1989,Abram:1986,Shih:1988}, remains a central tool for a wide and growing variety of quantum applications.

The enduring interest in the HOM effect proceeds from its role as a deterministic method for generating and manipulating two-photon entanglement, a valuable resource for quantum information processing \cite{EPR:1935,Ekert:1991,Dowling:2008,Giovannetti:2011}, especially quantum computing \cite{Kok:2007}. Owing to the measurement induced non-linearity resulting from post-selective techniques applied to the HOM output, it plays a crucial role in the operation of probabilistic universal gates such as the Controlled-Z (CZ) \cite{Kok:2007,Bouchard_2021} or the Controlled-NOT (CNOT) \cite{KLM:2001,Alsing_Hach:2019,Okamoto:2011}.

The HOM effect is important to the implementation of many specific applications in quantum computation relying upon single algorithm devices. Fisher et al. have shown computation on encrypted data \cite{Fisher:2014}, Harrow et al. and Cai et al. use quantum algorithms to solve systems of linear equations \cite{Harrow:2009,Cai:2013}, the computation of discrete and fractional Fourier transforms has been shown by Weimann et al. \cite{Weimann:2016}, and Humphreys et al. have shown linear optical quantum computing on a single spatial mode \cite{Humphreys:2013}. 

The HOM interference visibility is useful in evaluating the level of indistinguishability of photons from a variety of single-photon sources. Kaltenbaek et al. and Mosley et al. used this technique for Spontaneous Parametric Down Conversion (SPDC) sources \cite{Kaltenbaek:2006,Mosley:2008}. Numerous groups have used this method to investigate photons produced by quantum dots \cite{Sanaka:2009,Flagg:2010,Patel:2010,Wei:2014,Senellart:2017}, atomic vapors \cite{Felinto:2006,Chaneliere:2007,Yuan:2007,Yuan:2008,Chen:2008}, nitrogen-vacancy centers in diamond \cite{Bernien:2012,Sipahigil:2012,Sipahigil:2014}, molecules \cite{Kiraz:2005,Lettow:2010}, trapped neutral atoms \cite{Beugnon:2006,Specht:2011}, and trapped ions \cite{Maunz:2007}. 

The HOM effect is widely applied in quantum communication and quantum cryptography \cite{Gisin:2002,Pirandola:2019}. Commercial QKD systems have been shown experimentally to be vulnerable to several side channel attacks \cite{Lydersen:2010,Gerhardt:2011}. One proposed solution to overcome these security issues is a full-device-independent QKD scheme \cite{Mayers:1998,Acin:2007}, but its requirements are difficult to meet and it yields low secret key rates which is why the measurement-device-independent protocol was developed \cite{Braunstein:2012,Lo:2012}. The measurement-device-independent QKD protocol involves making a Bell state measurement by making photons arrive simultaneously on a beam splitter and observing the interference to establish the keys \cite{Bouchard_2021,Braunstein:2012,Lo:2012}. Liu et al. have demonstrated this measurement-device-independent protocol experimentally using time-bin phase-encoding \cite{Liu:2013}. For example, the passive round-robin differential phase-shift QKD protocol discussed by Guan et al. also capitalizes on the HOM effect \cite{Guan:2015}. Common schemes in quantum communication and quantum cryptography, such as quantum teleportation and entanglement swapping rely on the HOM induced entanglement \cite{Hofmann:2012,Narla:2016}. Quantum repeaters can be used in tandem with Bell state analyzers to entangle pairs of atoms which can then be used for these QKD protocols \cite{Sangouard:2011}. QKD appears to allow the exchange of secret information that cannot be intercepted, but security loopholes have been found in practical implementations \cite{Fung:2007,Zhao:2008}. 

Fabrication defects in large integrated devices containing multiple sequential devices such as Mach-Zehnder meshes remain an outstanding problem \cite{Miller:15, Hammerly:22a, Hammerly:22b}.  We compare our devices to a directional coupler tolerance of $5\%$ \cite{Mower:15}.  This is based on 193nm UV lithography standard \cite{Mikkelsen:14}, though many fab shops use older processes.  This can result in a splitting ratio varying by several percent.  Our devices show better tolerances to fabrication errors.

Here, we are motivated by the global perspective that theoretical proposals for enhancing the parametric flexibility of photonic HOM elements will directly impact many, if not all, applications relying on the HOM Effect in such an architecture. One example of this sort of enhancement has been presented earlier by the present authors in the form of higher-dimensional Hong-Ou-Mandel Manifolds (HOMM) within the parameter space of a double-bus Micro-Ring Resonator (db-MRR) \cite{Hach:2014}. 

The MRR can be thought of taking a beam splitter (one parameter), and expanding it into (in general) three parameters (a “top” and “bottom” directional coupler, and an internal phase angle), which are coupled together to form higher dimensional manifolds in which the HOM destructive interference condition can be met (vs the single point solution associated with a beam splitter) \cite{Hach:2014,Alsing_Hach:2017a} (see Appendix \ref{app:C} for a more complete discussion of this point). 
We use the term Hong-Ou-Mandel manifold (HOMM) to describe these higher dimensional solution sets in parameter space.  The extension to linear chains of MRRs allows for further design optimization to consider coupling and phase stability. MRRs also provide increased scalability and tunability over a simple beam splitter.

We examine significant enhancements to the HOMM that arise as a result of fabricating a relatively simple photonic circuit comprised of a serial chain of identical, balanced MRRs, and we propose in connection with them three figures of merit informing the design, fabrication, and characterization of any such structure in the context of a wide variety of applications. 

The outline of the paper is as follows: 
in Section II we define our approach to characterizing the HOM effect in an MRR, present the exact analytic solution of the one-dimensional HOMM for any number of MRRs in series, and define our first figure of merit $\tau_c$, the minimum MRR coupling value to achieve the HOM effect. 
In Section III we investigate the phase stability of our one-dimensional HOMM and define our second figure of merit $\bar{\xi}_N$, the curvature of the 1-dimensional HOMM. 
in Section IV, we then investigate the coupling stability of our one-dimensional HOMM and define our last figure of merit $\delta\tau_{N}$, the 5\% tolerance in the probability for complete destructive interference to occur on the HOMM. Finally in Section V we summarize our results and describe how these three figures of merit can be used in combination to inform design and fabrication. 
We also include in Appendix A, a review of the evanescent couplers used in this db-MRR and in Appendix B, a more detailed derivation of our main result, the exact analytic expression of the one-dimensional HOMM for any number of MRRs in series.

\section{MRR HOMM results}

\begin{figure}[H]
	\centering
	\includegraphics[width=\linewidth,keepaspectratio]{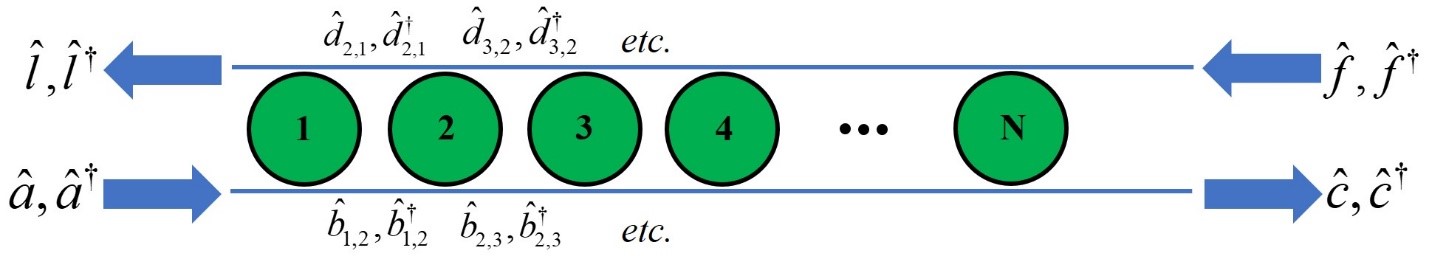}
	\caption{A serial chain of identically coupled, identical double-bus MRR; the input modes are \textit{a} and \textit{f}, and the output modes are \textit{c} and \textit{l}. The internal modes, \textit{b} and \textit{d}, enter into the derivation of the HOM Manifolds, but they do not enter explicitly into the results.}
	\label{fig:Nmrrsdiagram}
\end{figure}

Fig. \ref{fig:Nmrrsdiagram} shows a schematic representation of such a chain, indicating the notation that we adopt. We define the chain length of the circuit as \textit{N}, the number of identical MRRs in the chain. 
The MRRs are identical, each having a common round-trip phase shift $\theta_j\xrightarrow{}\theta$,  
equal transmission and reflection coefficients 
$(\tau_j,\kappa_j,\gamma_j,\eta_j)\xrightarrow{}(\tau,\kappa,\gamma,\eta)$,  
and are balanced, i.e.
$\kappa=\gamma$ and $\tau=\eta$. 
%
We invoke here the labeling scheme introduced in \cite{Hach:2014} for the coupling parameters; these are reviewed in Appendix A.
Further we adopt the phase convention whereby:

\begin{equation}
    \begin{split}
        \tau&=\tau^*=|\tau|\\
        \kappa&=i\sqrt{1-|\tau|^2}=i\sqrt{1-\tau^2}
    \end{split}
    \label{taukappaeqn}
\end{equation}

The phases indicated in Eqn.(\ref{taukappaeqn}) ensure the satisfaction of the reciprocity relations for each directional coupler \cite{Yariv:2000}, and they allow us to characterize the linear chain using the three real parameters, $(\tau,\theta,N)$ (see Appendix A). The one-dimensional HOMM of a chain having length $N$ is then given by the curve $\tau_N(\theta)$ (see Appendix B). Our prior work in \cite{Hach:2014} then serves as the $N=1$ basis for comparison and characterization of our new results for arbitrary chain lengths.
%

Owing to the linearity of the medium (negligible absorption), the evanescent nature of the coupling (negligible scattering), and the low intensities of the photons (negligible photonic losses) we work here in the lossless regime. As we show below, for many applications chain lengths of ~2-3 are optimal, justifying the lossless approximation for short chains. We have previously presented analyses of losses in the ‘unit cell’ db-MRR
\cite{Alsing_Hach:2017a,Alsing_Hach:2017b1,Alsing_Hach:2017b2}. The lesson from that work was that parametric HOMM structures (see Fig.~\ref{fig6contours} below) are persistent within reasonable experimental tolerances even in the presence of realistic loss mechanisms. 

One way to formulate the one-dimensional HOMM for a serial chain having length $N>1$ is to use the result that we have previously derived based upon a Discrete Path Integral (DPI) for the $N=1$ case \cite{Hach:2014,Skaar:2004}, along with the relevant application of the ``mode-swap algebra'' that we have described previously in proposing a scalable CNOT gate \cite{Alsing_Hach:2019}. Fig.\ref{fig:Nmrrsdiagram} explicitly displays the operators representing as quantum interconnects along each of the waveguides between each MRR. These operators $\{\hat{d}_{j,i},\hat{b}_{i,j}\}$ must then be algebraically eliminated in order to produce the desired ``S-matrix'' relating the input and outputs for the chain as a whole.

Using this approach along with mathematical induction \cite{Hach:2014,Alsing_Hach:2019}, one can show (see Appendix B) analytically that the probability for output photon coincidence from the system represented in Fig. \ref{fig:Nmrrsdiagram} is given by:
\begin{equation}
    P_{1,1}(\tau,\theta;N)=\frac{(1-4\tau^{2N}+\tau^{4N}+2\tau^{2N}\cos{\theta})^2}{(1+\tau^{4N}-2\tau^{2N}\cos{\theta})^2}
    \label{P11analytic}
\end{equation}
The HOM constraint can be expressed using
\begin{equation}
    P_{1,1}(\tau,\theta;N)=0
    \label{eqn12}
\end{equation}
For a chain having length $N$, Eqn.(\ref{eqn12}) defines a one-dimensional HOMM, $\tau_N(\theta)$.  After some more algebra, we find that the physical one-dimensional HOMM for the serial chain having length $N$ is given in closed form by 
\begin{equation}
    \tau_N(\theta)=(2-\cos{\theta}-\sqrt{3-4\cos{\theta}+\cos^2{\theta}})^{\frac{1}{2N}}
    \label{eqn13main}
\end{equation}

Eqn.(\ref{eqn13main}) encodes the principal result of this work. We now use this result to motivate our proposals for three figures of merit characterizing the practical implementation of the HOM Effect in integrated nanophotonics.

\begin{figure}[H]
	\centering
	\includegraphics[width=0.9\linewidth,keepaspectratio]{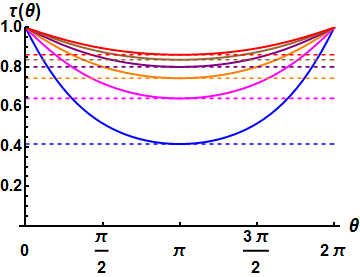}
	\caption{(color online): One dimensional HOM Manifolds (HOMM) for chains of identical MRRs having lengths N=1 (blue), 2 (magenta), 3 (orange), 4 (purple), 5 (brown), and 6 (red). The horizontal lines indicate the HOM cutoff value for the coupling parameter for each chain.}
	\label{fig:3}
\end{figure}

In Fig. \ref{fig:3} we plot the one-dimensional HOMM (1d-HOMM) for several chain lengths, including for comparison the $N=1$ case \cite{Hach:2014}. Several important features are apparent. First, the 1d-HOMM are all symmetric in $\tau$ about a round-trip phase shift of $\theta_0=\pi$. This phase shift turns out to be especially important in what follows, so we have labeled it with a subscript. 

Second, the cutoff coupling parameter below which the HOM Effect never occurs in the device increases monotonically with chain length. It is straightforward to determine this trend analytically,

\begin{equation}
    \tau_{c}=(3-\sqrt{8})^{\frac{1}{2N}}
\end{equation}

as shown by the horizontal lines in Fig. \ref{fig:3}. The HOM Effect is clearly attainable over the experimentally feasible range of couplings $0.5\leq\tau\leq0.9$, easily bracketing the important case of the 3dB coupler, $\tau_{3dB}=\frac{1}{\sqrt{2}}$.

\section{HOM Phase Stability}
It is apparent from Fig. \ref{fig:3} that the 1d-HOMM is particularly flat for any length chain near the point $(\tau_{c},\theta_0=\pi)$ , which we shall refer to as the optimal operating point for implementation of the HOM Effect. This flatness suggests a high degree of parametric stability in the HOM effect from a linear chain operating near this point compared to a directional coupler \cite{Mikkelsen:14}. Stability of this nature in concert with the inherent tunability of the silicon nanophotonic architecture implies a potentially large advantage for deploying HOM-based elements on scalable platforms. 

To make this more precise, we expand the coincidence probability given by Eqn.(\ref{P11analytic}) in $\tau$ and $\theta$ to lowest non-vanishing order about the optimal operating point,
\begin{equation}
    P_{1,1}(\tau,\theta;N)\approx\frac{1}{2}\frac{\partial^2P_{1,1}}{\partial\tau^2}\Bigr|_{\substack{0}}(\tau-\tau_{c})^2+\frac{1}{24}\frac{\partial^4P_{1,1}}{\partial\theta^4}\Bigr|_{\substack{0}}(\theta-\pi)^4
    \label{eq15}
\end{equation}
Evaluation of the non-vanishing expansion coefficients appearing in Eqn.(\ref{eq15}) is straightforward but not enlightening. Instead, the important features that emerge are that the HOM Effect near the optimal operating point is stable to second order in the coupling parameter and to fourth order in the round-trip phase shift. The HOM stability in $\theta$ is especially striking; It is evident that linear chains of identical db-MRRs share a universal third-order HOM stability when operated near the cutoff coupling, making the structure especially robust there against parametric drift due to environmental conditions. Motivated by this we identify the cutoff value of the directional coupling parameter, $\tau_{c}$, itself as one figure of merit characterizing the HOM performance of this class of integrated photonic structure. 

Once again referring to Fig. \ref{fig:3}, we can extend the notion of local HOM parametric stability near the optimal operating point to the global concept of HOM device stability. The round-trip phase shift for a photon having angular frequency $\omega$ propagating through an MRR having radius $R$ is given by 
\begin{equation}
    \theta=2\pi\frac{n(\omega)R}{c}\omega
\end{equation}
where $n(\omega)$ is the linear index of refraction of silicon. If we imagine operating conditions for a chain having length $N$ wherein the HOM Effect occurs with unit fidelity for a given photon angular frequency, $\omega$, and assuming that the coupling is independent of small changes in the frequency, $\delta\omega$, we see that the flatter the HOMM, $\tau_N(\theta(\omega))$, the larger the residual HOM fidelity over the support of the frequency spectrum for input photons. A shift $\delta\omega$ in photon frequency induces a round-trip phase shift of 
\begin{equation}
    \delta\theta=2\pi\frac{n(\omega)R}{c}\delta\omega
    \label{eq17}
\end{equation}
where we have assumed a flat dispersion relation for silicon over the relevant range of angular frequencies. Expanding the photon coincidence probability in MRR round-trip phase shift to lowest non-vanishing order about any given point on the 1d-HOMM for the chain gives
\begin{equation}
    P_{1,1}(\tau_N(\theta),\theta+\delta\theta;N)\approx\frac{1}{4}\cot^2{\left(\frac{\theta}{2}\right)}(\delta\theta)^2
    \label{eqn18}
\end{equation}
where we have explicitly indicated that the value of the coupling parameter is the one appropriate to the point on the HOMM about which we are expanding in $\theta$. Note that Eqn.(\ref{eqn18}) is consistent with Eqn.(\ref{eq15}) for $\theta\xrightarrow{}\pi$; the second order correction in $\theta$ vanishes there. Eqn.(\ref{eqn18}) shows that the HOM stability with respect to the MRR round-trip phase shift is maintained to first order along the entire 1d-HOMM. Further, the fact that the second-order correction term given in Eqn.(\ref{eqn18}) is independent of the chain length implies that the local round-trip phase HOM stability at each point on the 1d-HOMM is universal to this class of device. In view of Eqn.(\ref{eq17}), this results in a very high-fidelity HOM output from any given serial chain over a non-vanishing range of input photon frequencies. The general design lesson in this is that serial chains with flatter 1d-HOMMs are better suited to accommodate practical achievement of the HOM Effect for input photons with finite bandwidths. It is in this sense that the db-MRR structure, even for $N=1$, acts as a sort of filter for the HOM Effect itself. Whereas HOM Fidelity disappears at a beam splitter rapidly when the input photons become non-monochromatic, the deterioration is much less violent in the MRR-based structure. This is a distinct advantage of this architecture that relies directly on the existence of the HOMM. 

Motivated by the universal first-(at least)-order HOM stability near the 1d-HOMM and the independence from chain length of the lowest order correction term, we propose a global figure of merit to characterize the ‘overall flatness’ of the HOMM for a chain having length $N$. We characterize the HOM round-trip phase stability of such a device using the average curvature of the 1d-HOMM
\cite{Millman_Parker:1977} 
\begin{equation}
    \bar{\xi}_N\equiv\frac{1}{2\pi}\int_0^{2\pi}\frac{|\frac{d^2\tau_N(\theta)}{d\theta^2}|}{(1+\frac{d\tau_N(\theta)}{d\theta})^{\frac{3}{2}}}\,d\theta
\end{equation}
Smaller values of $\bar{\xi}_N$ characterize serial chain HOM structures with higher phase stability. Using Eqn.(\ref{eqn13main}), one can, in principle, develop a complicated analytical form for $\bar{\xi}_N$; instead we take the pragmatic approach of computing numerically the average curvature for the 1d-HOMM in linear chains having lengths of $N=1,...,10$. These results are displayed below in Table I. The monotonic trend toward overall flatness is apparent suggesting that longer chains offer increasing HOM phase stability, which, in view of Eqn.(\ref{eq17}) suggests persistence in the HOM visibility over broader spectral ranges of input photons. 

\section{HOM Coupling Stability}
We now investigate HOM stability against small changes in the coupling parameter, $\tau$, about operating points on the 1d-HOMM, especially near the optimal operating point, $(\tau_{c},\theta_0=\pi)$. It is easy to show that the lowest order correction term is of the second order in $\tau$ for operation of the linear chain near \textit{any} point on the 1d-HOMM, 
\begin{equation}
    P_{1,1}(\tau,\theta;N)\approx\frac{1}{2}\frac{\partial^2P_{1,1}}{\partial\tau^2}\Bigr|_{\substack{HOMM}}(\tau-\tau_{N}(\theta))^2
    \label{eqn20}
\end{equation}
Near the optimal operating point, this result assumes the form
\begin{equation}
    P_{1,1}(\tau,\theta_0;N)\approx\frac{2N^2(\tau-\tau_{c})^2}{\tau_{c}^2}
\end{equation}
One can use Eqn.(\ref{P11analytic}) and (\ref{eqn13main}) to arrive at a closed form for the general, second-order expansion coefficient in $\tau$, as indicated in Eqn.(\ref{eqn20}), but this form is cumbersome, so we omit here. 

\begin{figure}[H]
	\centering
        \includegraphics[width=0.475\linewidth,keepaspectratio]{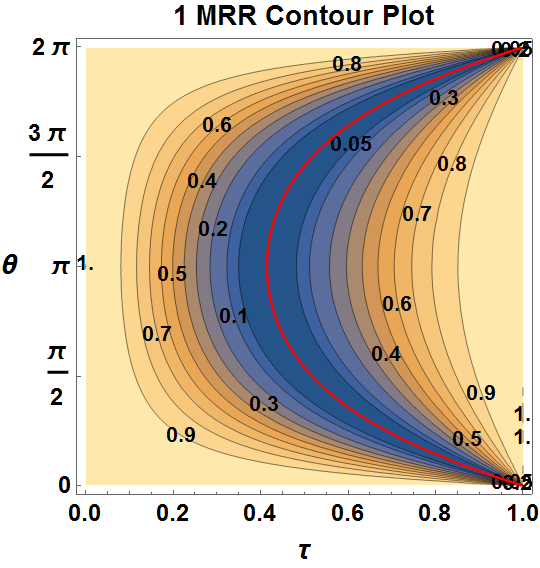}
        \includegraphics[width=0.475\linewidth,keepaspectratio]{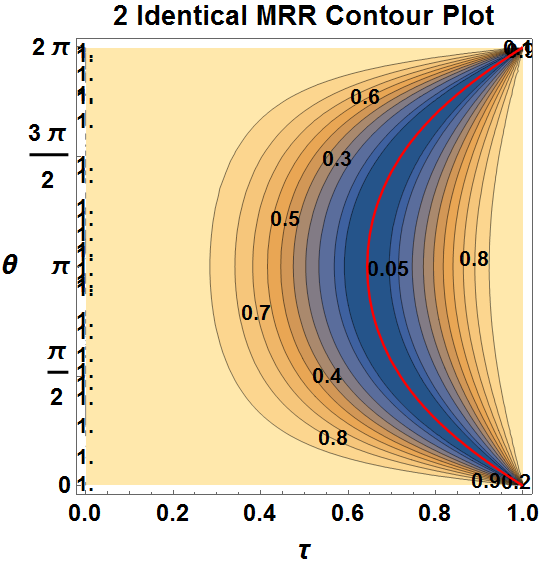}
        \includegraphics[width=0.475\linewidth,keepaspectratio]{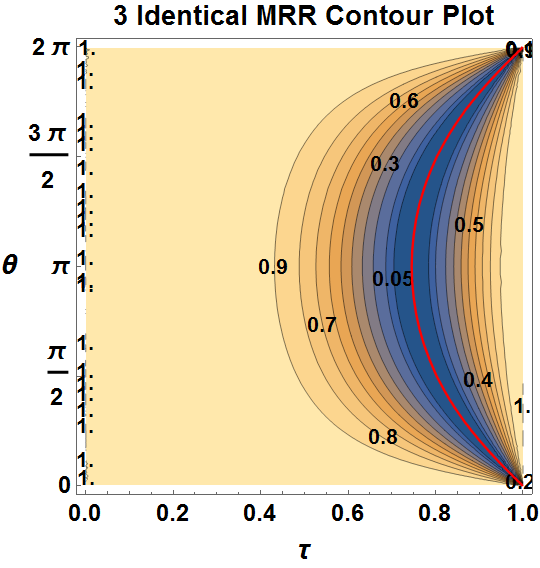}
    \caption{Contour plots of $P_{1,1}(\tau,\theta;N)$ vs. $\tau$ and $\theta$ for chains having lengths (top left) $N=1$, (top right) $N=2$, and (bottom) $N=3$. The red curve in each contour plot is the exact curve of zero probability, described in Eqn.(\ref{eqn13main}). These crescent shaped plots are qualitatively similar for chains of all lengths.}
    \label{fig6contours}
\end{figure}

In order to characterize the HOM coupling stability for a linear chain, we consider in Fig.\ref{fig6contours} contour plots of the output photon coincidence probability versus the coupling parameter and the round-trip phase shift for chains having lengths of 1, 2, and 3. The basic `crescent' shape prevalent in Fig.\ref{fig6contours} is characteristic of chains of all lengths. The nadir of the valley running though the deepest crescent, shown as a solid red curve on each contour plot, is the 1d-HOMM described in Eqn.(\ref{eqn13main}). The HOM effect occurs at every point along this red curve.

We pay particular attention to the relatively wide crescent shaped region between the $P_{1,1}=0.05$ level curves in each part of Fig.\ref{fig6contours}. These are the parametric regions in which the HOM Effect is obtained with a fidelity of 95\% of greater, a level typical of careful experimental measurements \cite{Alsing_Hach:2017a}. The remarkable HOM stability of the chain is evident in the extent of this high-fidelity region, especially near the optimal operating point. We use this observation as the basis for a figure of merit characterizing the HOM stability in the coupling parameter. 

We define the practical HOM-tolerance level for a linear chain having length \textit{N} to be the probability of an undesired output photon coincidence, 
\begin{equation}
    \epsilon_N\equiv P_{1,1}
    \label{eq22}
\end{equation}
where it is understood that Eqn.(\ref{eq22}) only has meaning in the neighborhood of an exact HOMM, where $P_{1,1}<<1$.

For a given tolerance, $\epsilon_N$, the level curves delineating the upper (+) and lower (-) bounds of the HOM region can be computed in exact form to be:
\begin{widetext}
\begin{equation}
    \tau_N^{(\pm)}(\theta)=\frac{1}{(1\mp\sqrt{\epsilon_N})^{\frac{1}{2N}}}\left\{2-(1\pm\sqrt{\epsilon_N})\cos{\theta}-\sqrt{4-(1\mp\sqrt{\epsilon_N})^2-4(1\pm\sqrt{\epsilon_N})\cos{\theta}+(1\pm\sqrt{\epsilon_N})^2\cos^2\theta}\right\}^{\frac{1}{2N}}
\end{equation}
\end{widetext}
which, at its widest near $\theta_0=\pi$ becomes 
\begin{equation}
    \tau_N^{(\pm)}(\pi)=\left[\frac{3-2\sqrt{2}\sqrt{(1\pm\sqrt{\epsilon_N})}\pm\sqrt{\epsilon_N}}{1\mp\sqrt{\epsilon_N}}\right]^{\frac{1}{2N}}
    \label{eqn24}
\end{equation}
In order to characterize the general HOM coupling stability within experimental tolerance $\epsilon_N$ for a linear chain having length $N$, we propose a figure of merit defined by 
\begin{equation}
    \delta\tau_N(\epsilon_N)\equiv|\tau_N^{(+)}(\pi)-\tau_N^{(-)}(\pi)|
    \label{eqn25}
\end{equation}

We choose this metric motivated by the reasonable assumption that linear chains deployed as HOM-based elements within photonic circuits will be integrated to work near the optimal operating point, where, as we have shown above, a high degree of HOM stability against small design, environmental, and spectral fluctuations can be realized. 

\section{Conclusion}
We summarize results for several chain lengths in Table I. Our approach in this work has been to introduce useful design parameters for the rather idealized case of identical MRRs, allowing closed form results, but with the idea that the same set of metrics, computed numerically, will be useful for designing and characterizing the HOM responses from complex MRR-based structures useful for a wide range of scientific and engineering applications.

With Table I in view, we now summarize the utility and importance of the results we have presented in this work. 
It is clear that, as we progress to longer chains of identical db-MRRs, there is a three-way competition between the spectral stability (via $\bar{\xi}_N$), the coupling stability (via $\delta\tau_{N}$), and the minimum coupling, $\tau_{c}$, for which the HOM can be realized. As one considers longer chains, the spectral stability increases ($\bar{\xi}_N$ decreases), the coupling stability for any given NOON state fidelity decreases ($\delta\tau_{N}$ decreases), and the coupling cutoff creeps toward larger values of transmission for each MRR in the chain.
\begin{table}[H]
\caption{Table summarizing characteristics for series of up to ten identical rings. In this table $\delta\tau_{N}$ uses a value of $\epsilon_N=0.05$.}
\begin{center}
\begin{tabular}{|c|c|c|c|}
 \hline
 \;\;\;N\;\;\; & \;\;\;$\tau_{c}$\;\;\; & \;\;\;$\bar{\xi}_N$\;\;\; & \;\;\;$\delta\tau_{N}$\;\;\; \\  [0.25ex]
 \hline
  \;\;\;1\;\;\; &  \;\;\;0.4142\;\;\; & \;\;\;0.1424\;\;\; & \;\;\;0.1323\;\;\;  \\   [0.5ex]
  2 & 0.6436 & 0.0772 & 0.1029 \\ [0.5ex]
 3 & 0.7454 & 0.0523 & 0.0796 \\ [0.5ex]
 4 & 0.8022 & 0.0395 & 0.0642 \\ [0.5ex]
 5 & 0.8384 & 0.0317 & 0.0537 \\ [0.5ex]
 6 & 0.8634 & 0.0264 & 0.0461 \\ [0.5ex]
 7 & 0.8817 & 0.0227 & 0.0404 \\ [0.5ex]
 8 & 0.8957 & 0.0199 & 0.0359 \\ [0.5ex]
 9 & 0.9067 & 0.0177 & 0.0323 \\ [0.5ex]
 10 & 0.9156 & 0.0159 & 0.0294 \\ [0.5ex]
 \hline
\end{tabular}
\end{center}
\end{table}

The HOM effect cannot be obtained at any $\tau$ value less than $\tau_c$ for a given number of MRRs. There is nothing inherently desirable or undesirable about any particular value of $\tau_c$ (since any value is achievable, in theory, due to the extreme tunability of the device), but one must keep in mind that, for example, if using a linear array with 5 MRRs in it the minimum $\tau$ value, $\tau_c$, that can achieve the HOM effect is $\tau_c\approx0.84$ as shown in Table 1. If such a high $\tau_c$ is unrealistic to maintain for a specific implementation in a device, a linear chain of fewer MRRs should be considered.  That is the “merit” in knowing the $\tau_c$ value for each chain length, N. As long as higher $\tau$ coupling values can be achieved, the design choice is between coupling stability and spectral stability. Each successive MRR added to the linear chain decreases the coupling stability and increases the spectral stability of the HOMM where the HOM effect is achieved.

In view of the many applications for the HOM effect that we have reviewed above, it is easy to imagine various optimizations of this competition for maximizing the efficiency of any HOM implementation within a quantum photonic circuit. For example, for a chain that is to be used near the 3dB (50/50) coupling point, one is limited to having, at most, 2 MRRs in the chain. Within that restrictive engineering constraint, however there is still some design freedom. For a system in which the coupling parameter is very stable and well-controlled, it should be preferable to use two series MRRs for the increase in phase (i.e. spectral) stability they offer in comparison with a similarly coupled, individual MRR.

In this work, we have presented three figures of merit that characterize important design and optimization opportunities for a class of scalable structures that can readily be integrated into quantum photonic circuits using fabrication techniques that are already implemented routinely. Given the long and persistent history of the HOM effect as an essential tool for quantum photonics, the work have presented here has the potential to inform the design, optimization, and implementation of an extremely wide range of photonic networks central to next generation quantum information processing, sensing, and metrology.

\begin{acknowledgments}
E.E.H. and J.M. acknowledge support from the United States Air Force Research Laboratory (AFRL) Summer Faculty Fellowship Program for providing support for this work. P.M.A. and A.M.S. acknowledges support from the Air Force Office of Scientific Research (AFOSR). Any opinions, findings and conclusions or recommendations expressed in this material are those of the author(s) and do not necessarily reflect the views of the Air Force Research Laboratory (AFRL).
\end{acknowledgments}

\clearpage
\newpage
\appendix

\section{Review of the db-MRR: Evanescent coupling parameters}

\begin{figure}[H]
	\centering
	\includegraphics[width=\linewidth,keepaspectratio]{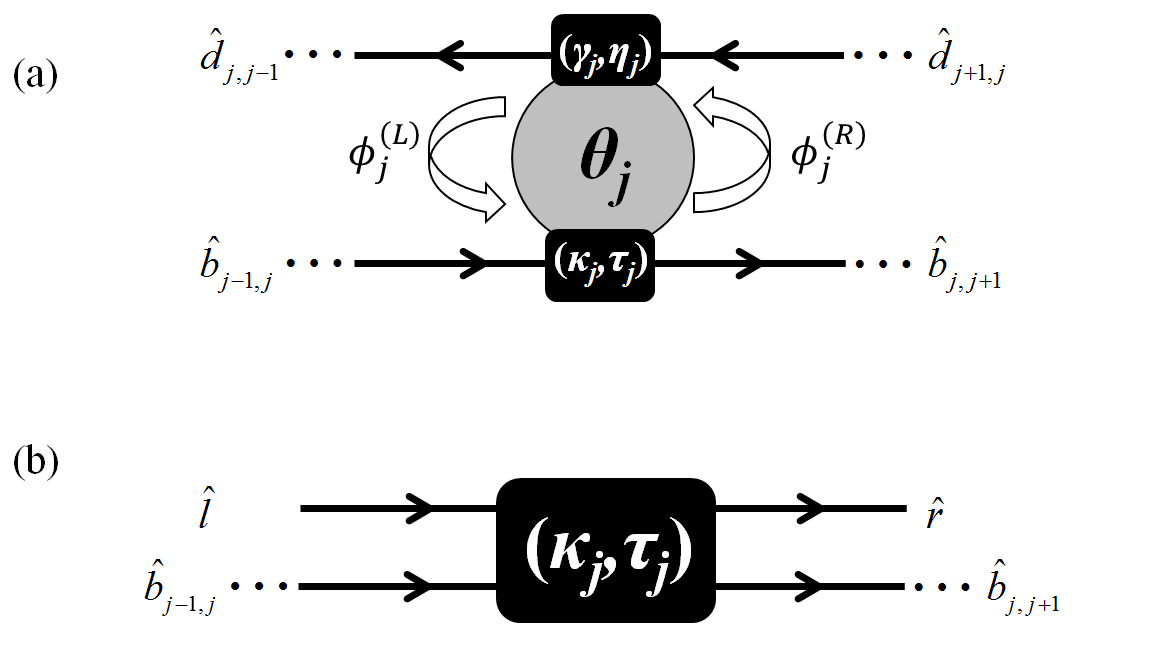}
	\caption{(a) Schematic diagram of the $j^{th}$ `unit cell' for a serial chain of db-MRRs. This version explicitly displays the general couplings to the waveguides and allows us to re-introduce the parameters we use throughout the body of the work. The directional input/output nature of the device is indicated by the arrows. (b) A close-up view of one of the evanescent couplers, internal MRR mode operators, $\hat{l}$ and $\hat{r}$, are included to emphasize the input/output ports of the coupler}
	\label{supmatfig1}
\end{figure}

To provide explicit context for the parameters we use throughout the work, we show in Fig. \ref{supmatfig1} schematic diagrams for the coupling of (a) a `unit cell' of the serial chain and (b) one of the directional couplers within the unit cell in the general case. 

The directional coupler in Fig. 1b is described by the linear mode operator transformation

\begin{equation}
    \begin{pmatrix}
    \hat{b}_{j,j+1} \\
    \hat{r}
    \end{pmatrix}
    =
    \begin{pmatrix}
    \tau_j & \kappa_j \\
    -\kappa_j^* & \tau_j^*
    \end{pmatrix}
    \begin{pmatrix}
    \hat{b}_{j-1,j} \\
    \hat{l}
    \end{pmatrix}
\end{equation}
where, using the phase convention we have adopted for the coupling parameters, the bosonic commutation relations are enforced by the usual reciprocity relation

\begin{equation}
    |\kappa_j|^2+|\tau_j|^2=1
\end{equation}
with analogous relationships describing the coupling at the top of the MRR shown in Fig. \ref{supmatfig1}. Using either the Discrete Path Integral (DPI) approach we have developed in earlier work \cite{Skaar:2004,Hach:2014} or the boundary value approach \cite{Raymer:2013,Alsing_Hach:2017a}, it is straightforward to show that the operator input/output relations appropriate for the Heisenberg picture description of photon transport through the db-MRR shown in Fig. \ref{supmatfig1} are given by the linear system

\begin{equation}
    \begin{pmatrix}
    \hat{b}_{j,j+1}^\dagger \\[6pt]
    \hat{d}_{j,j-1}^\dagger
    \end{pmatrix}
    =
    \begin{pmatrix}
    T_{11}^{(j)} & T_{12}^{(j)} \\[6pt]
    T_{21}^{(j)} & T_{22}^{(j)}
    \end{pmatrix}
    \begin{pmatrix}
    \hat{b}_{j-1,j}^\dagger \\[6pt]
    \hat{d}_{j+1,j}^\dagger
    \end{pmatrix}
    \label{supeqn3}
\end{equation}
where
\begin{equation}
\begin{aligned}
    T_{11}^{(j)}&=\frac{\eta_j^*-\tau_je^{i\theta_j}}{\eta_j^*\tau_j^*-e^{i\theta_j}} & T_{12}^{(j)}&=\frac{\gamma_j\kappa_j^*e^{i\phi_j^{(L)}}}{\eta_j^*\tau_j^*-e^{i\theta_j}} \\
    T_{21}^{(j)}&=\frac{\kappa_j\gamma_j^*e^{i\phi_j^{(R)}}}{\eta_j^*\tau_j^*-e^{i\theta_j}} & T_{22}^{(j)}&=\frac{\tau_j^*-\eta_je^{i\theta_j}}{\eta_j^*\tau_j^*-e^{i\theta_j}}
\end{aligned}
\label{supeqn4}
\end{equation}
The phase shifts that appear in Eqn.(\ref{supeqn4}) are related via $\phi_j^{(R)}+\phi_j^{(L)}=\theta_j$, where $\theta_j$ is the round-trip phase shift for the $j^{th}$ MRR in the chain.

We re-write Eqns.(\ref{supeqn3}) and (\ref{supeqn4}) in the form

\begin{equation}
    \begin{pmatrix}
    \hat{b}^\dagger_{j,j+1} \\[6pt]
    \hat{d}^\dagger_{j,j-1}
    \end{pmatrix}
    =
    \mathbf{T}^{(j)}
    \begin{pmatrix}
    \hat{b}^\dagger_{j-1,j} \\[6pt]
    \hat{d}^\dagger_{j+1,j}
    \end{pmatrix}
\end{equation}

In this work we treat the case in which all of the MRRs are identical $(\theta_j\xrightarrow[]{}\theta)$ and balanced $(\phi_j^{(R)}=\phi_j^{(L)}\xrightarrow[]{}\frac{\theta_j}{2})$. We further assume that the MRRs are all symmetrically and identically coupled,
\begin{equation}
    (\tau_j=\eta_j,\kappa_j=\gamma_j)\xrightarrow[]{}(\tau,\kappa)
\end{equation}
Finally, without further loss of generality we take the direct transmission coefficients for each of the directional couplers to be real, resulting in Eqn.(1) within the main text. With these modifications the photonic transfer matrix for the $j^{th}$ MRR assumes the form:
\begin{equation}
    \mathbf{T}^{(j)}\xrightarrow{}\frac{\tau^2e^{i\frac{\theta}{2}}-e^{-i\frac{\theta}{2}}}{1+\tau^4-2\tau^2\cos{\frac{\theta}{2}}}
    \begin{pmatrix}
    -2i\tau\sin{\frac{\theta}{2}} & 1-\tau^2 \\
    1-\tau^2 & -2i\tau\sin{\frac{\theta}{2}}
    \end{pmatrix}
\end{equation}

\begin{figure}[H]
	\centering
	\includegraphics[width=\linewidth,keepaspectratio]{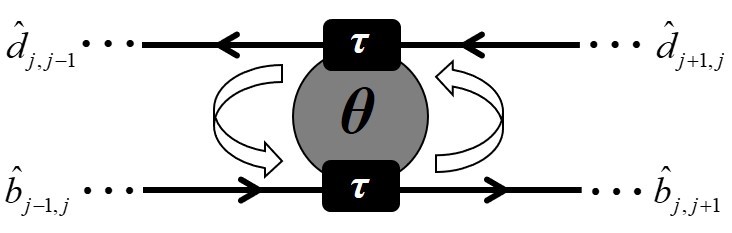}
	\caption{The $j^{th}$ db-MRR in a serial chain having arbitrary length N. As described in the body of the paper, we have assumed that all of the symmetric, balanced MRRs are identical and that they are identically coupled to the waveguides. We represent each coupling in the figure using the real transmission coefficient $\tau$ for the directional couplers involved.}
	\label{fig:2}
\end{figure}

Fig. \ref{fig:2} represents the ``unit'' cell of the system we consider after these parametric choices have been imposed.

\section{Derivation of HOM condition}

In order to analyze the photonic transport properties of non-trivial chains $(N>1)$, we must apply the combination rule for $\mathbf{T}$ matrices that properly relates the creation operators describing the input modes to those describing the output modes in accordance with our Heisenberg Picture description of the interaction. Referring to the overall transformation matrix for the creation operators due to an MRR chain of length $N$ as $\mathbf{M}_N$ such that
\begin{equation}
    \begin{pmatrix}
    \hat{a}^\dagger \\
    \hat{f}^\dagger
    \end{pmatrix}
    =
    \mathbf{M}_N
    \begin{pmatrix}
    \hat{c}^\dagger \\
    \hat{l}^\dagger
    \end{pmatrix}
    \label{eq4}
\end{equation}
Applying the mode-swap algebra that we introduced in \cite{NLPSG} suitably tailored to the present $2\times2$ case, we express the formal solution for $\mathbf{M}_N$ using
\begin{equation}
\mathbf{M}_N=\mathcal{S}_{2}^{(2)}
	\left[
		\prod\limits_{j=1}^N
	\mathcal{S}_{2}^{(2)}
	\left[
	\mathbf{T}^{(j)}
	\right]
	\right]
	\label{MNeqn}
\end{equation}
In writing Eqn.(\ref{MNeqn}) we have introduced notation whereby $\mathcal{S}_{m}^{(k)}$ operates on the square matrix $\mathbf{A}$ having dimension \textit{k} in such a way as to produce another square matrix having dimension \textit{k} appropriate to the linear system with the dependent variable in row \textit{m} `swapped' with the independent variable from that same row, for example:
\begin{equation}
    \begin{split}
        \text{If} \begin{pmatrix}
        x' \\
        y'
        \end{pmatrix}
        &=
        \begin{pmatrix}
        a & b \\
        c & d
        \end{pmatrix}
        \begin{pmatrix}
        x \\
        y
        \end{pmatrix}
        \\
        \text{Then} \begin{pmatrix}
        x' \\
        y
        \end{pmatrix}
        &=
        \mathcal{S}_{2}^{(2)}
        \left[
        \begin{pmatrix}
        a & b \\
        c & d
        \end{pmatrix}
        \right]
        \begin{pmatrix}
        x \\
        y'
        \end{pmatrix} \\
        &{} \\
        \\
    \end{split}
\end{equation}

Working in the lossless, continuous-wave (cw) approximation, we derive analytically the one-dimensional HOMM in the serial chain of identical MRRs. The input state to the chain is taken to be 
\begin{equation}
    \ket{in}=\ket{1}_a \otimes \ket{1}_f \equiv \ket{1_a,1_f}=\hat{a}^\dagger\hat{f}^\dagger\ket{vac}
\end{equation}
According to Eqn.(\ref{eq4}) the corresponding output state has the form:

\begin{widetext}
\begin{equation}
\begin{split}
    \ket{out}&=[(\mathbf{M}_N)_{11}\hat{c}^\dagger+(\mathbf{M}_N)_{12}\hat{l}^\dagger][(\mathbf{M}_N)_{21}\hat{c}^\dagger+(\mathbf{M}_N)_{22}\hat{l}^\dagger]\ket{vac} \\
    &=(\mathbf{M}_N)_{11}(\mathbf{M}_N)_{21}\ket{2_c,0_l}+(\mathbf{M}_N)_{12}(\mathbf{M}_N)_{22}\ket{0_c,2_l}+[(\mathbf{M}_N)_{11}(\mathbf{M}_N)_{22}+(\mathbf{M}_N)_{12}(\mathbf{M}_N)_{21}]\ket{1_c,1_l}
\end{split}
\label{longeq8}
\end{equation}
\end{widetext}

Making it obvious that the HOM Effect is obtained whenever the coefficient of the state $\ket{1_c,1_f}$ vanishes (the permanent of $\mathbf{M}_N)$). 
It is easy to show that the resulting output state is the well-known, properly normalized two-photon NOON state \cite{Dowling:2008}, 
\begin{equation}
    \ket{2::0;\Phi}=\frac{1}{\sqrt{2}}(\ket{2_c,0_l}+e^{i\Phi}\ket{0_c,2_l})
\end{equation}
To isolate the HOMM, we consider the output coincidence probability for photons in the state given in Eqn.(\ref{longeq8}),
\begin{equation}
\hspace{-0.25in}
    P_{1,1}(\tau,\theta;N)=|(\mathbf{M}_N)_{11}(\mathbf{M}_N)_{22}+(\mathbf{M}_N)_{12}(\mathbf{M}_N)_{21}|^2
\end{equation}

Using mathematical induction and after some algebraic manipulations, one can show analytically that the probability for output photon coincidence is given by:
\begin{equation}
    P_{1,1}(\tau,\theta;N)=\frac{(1-4\tau^{2N}+\tau^{4N}+2\tau^{2N}\cos{\theta})^2}{(1+\tau^{4N}-2\tau^{2N}\cos{\theta})^2}
    \label{P11appendix}
\end{equation}
The HOM constraint can be expressed using
\begin{equation}
    P_{1,1}(\tau,\theta;N)=0
    \label{eqn12appendix}
\end{equation}
For a chain having length $N$, Eqn.(\ref{eqn12appendix}) defines a one-dimensional HOMM, $\tau_N(\theta)$.  After some more algebra, we find that the physical one-dimensional HOMM for the serial chain having length $N$ is given in closed form by 
\begin{equation}
    \tau_N(\theta)=(2-\cos{\theta}-\sqrt{3-4\cos{\theta}+\cos^2{\theta}})^{\frac{1}{2N}}
    \label{eqn13appendix}
\end{equation}

\section{The relationship between the transfer matrices of  a conventional BS and a MRR.}\label{app:C}
In this appendix we show how the HOM effect in the MRR can be considered as a generalization of the HOM using a conventional  beam splitter.

We model the conventional beam splitter (BS) unitary transfer matrix (between input and output modes) with transmissivity  $\cos^2(\theta/2)$ by the $O(2)$ rotation matrix: 
\begin{equation}\label{BS:matrix}
 T_{BS}(\theta) =     
  \begin{pmatrix}
    \cos(\theta/2) & \sin(\theta/2) \\[6pt]
    -\sin(\theta/2) & \cos(\theta/2)
  \end{pmatrix}, 
  \begin{pmatrix}
    \cos(\theta/2) & i\,\sin(\theta/2) \\[6pt]
    i\,\sin(\theta/2) & \cos(\theta/2)
  \end{pmatrix},
 \end{equation}  
where the phases of the off-diagonal terms are either chosen as $\pm 1$ Eqn.(\ref{BS:matrix})(left), or as $i$ Eqn.(\ref{BS:matrix})(right). 
Here $0\le\theta\le\pi$ with a balanced 50:50 BS occuring at $\theta=\pi/2$. 
Note that the HOM effect occurs when the permanent of the matrix $T_{BS}$ is zero, i.e. $\cos^2(\theta/2)-\sin^2(\theta/2)=\cos(\theta)=0 \Rightarrow \theta=\pi/2$, the 50:50 BS condition. 

Now, a general two-port device (2 inputs, 2 outputs) can be modeled by an $SU(2)$ matrix, which we can always write in the form
\begin{equation}\label{MRR:matrix}
\hspace{-0.5in}
 T_{MRR} =     
  \begin{pmatrix}
    T_{11} & T_{12} \\[6pt]
    T_{21} & T_{22}
  \end{pmatrix} 
\to
  \begin{pmatrix}
    |T_{11}|\,e^{i\,\theta_{11}} & \sqrt{1-|T_{11}|^2}\,e^{i\,\theta_{12}} \\[6pt]
    -\sqrt{1-|T_{11}|^2}\,e^{i\,\theta_{21}} & |T_{11}|\,e^{i\,\theta_{22}}
  \end{pmatrix}
 \end{equation} 
 which automatically ensures the unitarity requirement that the inner product of each column (row) with itself is $1$. 
 Unitarity also requires that the inner product of a column (row) with a different column (row) yields $0$.
 Applying this to Eqn.(\ref{MRR:matrix}) yields the additional phase constraint: 
 
\begin{equation}\label{phase:constraint}
\theta_{11}+\theta_{22}=\theta_{12}+\theta_{21}.
\end{equation}

 An analysis of the HOM effect for an arbitrary dual Fock input state  $\ket{n,m}_{ab}$  (with $n$ and $m$ both odd) reveals that any arbitrary choice of phases satisfying the above phase constraint Eqn.(\ref{phase:constraint}), also yields the HOM effect (and in fact a more generalized version of it), for the complete destructive interference of the quantum amplitude for output coincidence counts \cite{eHOM_Alsing:2022}.

 Using Eqn.(\ref{MRR:matrix}), the ordinary BS Eqn.(\ref{BS:matrix}) is recovered by choosing 
 $T_{11}=\cos(\theta/2)$ (transmission amplitude) with $0\le\theta\le\pi$, and either the phase choices of 
 $\theta_{ij}=0$ for Eqn.(\ref{BS:matrix})(left) or 
 $\{\theta_{11}=\theta_{22}=0,\,\theta_{12}=\pi/2,\, \theta_{21}=-\pi/2\}$ for Eqn.(\ref{BS:matrix})(right).

For the MRR, we have from Eqn.(\ref{supeqn4})
\begin{equation}\label{T11:magnitude}
    |T_{11}^{(j)}|=\left|\frac{\eta_j^*-\tau_je^{i\theta_j}}{\eta_j^*\tau_j^*-e^{i\theta_j}}\right| \equiv \cos(\varphi/2), 
\end{equation}
where $\cos(\varphi)=\cos(\varphi(\eta_j,\tau_j,\theta_j))$ is a Fabry-Perot-like function of the upper ($\eta_j$), and lower ($\tau_j$) physical beam splitter transmission coefficients of the ($jth$) MRR, and $\theta_j$ is the round trip phase angle. 
But as discussed above, the particular values of the phases $\theta_{ij}$ of $T_{ij} = |T_{ij}|\,e^{i\,\theta_{ij}}$ do not effect HOM effect, as long as they satisfy the phase constraint Eqn.(\ref{phase:constraint}). (Note that in $T_{21}$ we have already factored out a crucial $-1$ sign in its definition).
Thus, if we for a moment chose the phases as $\theta_{ij}=0$, then $T_{MRR}$ would have the form of an \textit{effective} conventional BS: 
\begin{equation}\label{MRR:as:effective:BS:matrix}
T_{MRR}\overset{\theta_{ij}\to 0}{\longrightarrow}
\small{
\begin{pmatrix}
    \cos(\varphi/2) & \sin(\varphi/2) \\[6pt]
    -\sin(\varphi/2) & \cos(\varphi/2)
  \end{pmatrix}
}
\end{equation}
It is in the sense of Eqn.(\ref{MRR:as:effective:BS:matrix}) with defintion Eqn.(\ref{T11:magnitude}) that we mean that the MRR ``acts like'' an expansion of the conventional BS (with transmission coefficient $\cos(\theta/2)$), to two BS (with transmission coefficients $\eta_j$ and $\tau_j$), plus an internal phase angle $\theta_j$. 

Finally, for the MRR, the HOM effect also occurs when the permanent of the matrix $T_{MRR}$ is zero, i.e. $|T_{11}|^2\,e^{i(\theta_{11}+\theta_{22})} -
 (1-|T_{11}|^2)\,e^{i(\theta_{12}+\theta_{21})}=0 
 \Rightarrow 2|T_{11}|^2-1 = 2\cos(\varphi/2)^2-1 = \cos(\varphi)=0$, where we have used $|T_{11}|\equiv \cos(\varphi/2)$, and the phase constraint (to factor out the phases). Thus, for the MRR the HOM effect occurs for the effective BS angle condition $\varphi=\pi/2$, which by 
 Eqn.(\ref{T11:magnitude}) is a complicated function of the MRR upper and lower physical beam splitters and round trip phase angle. The fact that MRR HOM condition $\varphi(\eta_j,\tau_j,\theta_j)=\pi/2$ occurs in a larger parameter space is what gives rise to a manifold of HOM solutions (hence the term HOMM), vs the single point solution $\theta=\pi/2$ for the conventional BS.

%
\end{document}